\documentclass[aps,prl,twocolumn,superscriptaddress,showpacs,floatfix]{revtex4}
\usepackage{amsmath,graphicx}
\newcommand{\ket}[1]{\vert #1 \rangle}
\newcommand{\bra}[1]{\langle #1 \vert}
\newcommand{\SLM}{{\scriptstyle \!S\!L\!M}}
\newcommand{\HH}{{\hbox{\small HH}}}

\newcommand{\VV}{{\hbox{\small VV}}}
\begin{document}
\title{Programmable entanglement oscillations in a non Markovian channel}
\author{Simone Cialdi}\email{simone.cialdi@mi.infn.it}
\affiliation{Dipartimento di Fisica dell'Universit\`a degli Studi
di Milano, I-20133 Milano, Italia.}
\affiliation{INFN, Sezione di Milano, I-20133 Milano, Italia.}
\author{Davide Brivio}
\affiliation{Dipartimento di Fisica dell'Universit\`a degli Studi
di Milano, I-20133 Milano, Italia.}
\author{Enrico Tesio}
\affiliation{Dipartimento di Fisica dell'Universit\`a degli Studi
di Milano, I-20133 Milano, Italia.}
\author{Matteo G. A.~Paris}
\affiliation{Dipartimento di Fisica dell'Universit\`a degli Studi
di Milano, I-20133 Milano, Italia.}
\date{\today}
\begin{abstract}
We suggest and demonstrate an all-optical experimental setup to observe
and engineer entanglement oscillations of a pair of polarization qubits
in a non-Markovian channel. We generate entangled photon pairs 
by spontaneous parametric downconversion (SPDC), and then insert a 
programmable spatial light modulator in order to impose a polarization
dependent phase-shift on the spatial domain of the SPDC output and to
create an effective non-Markovian environment. Modulation of the
enviroment spectrum is obtained by inserting a spatial grating on the
signal arm.  In our experiment, programmable oscillations of
entanglement are achieved, with the maximally revived state that
violates Bell's inequality by $17$ standard deviations.
\end{abstract}
\maketitle
Entanglement of a bipartite system is usually degraded by the
interaction of each subsystem with the environment, which induces
decoherence, i.e. an irreversibile loss of information from the
system to the rest of the universe \cite{zur03,sch05}. If the
interaction is Markovian, i.e. the loss of information is
unidirectional, from the system to the environment, then entanglement
monotonically decreases and may be also destroyed in a finite time
\cite{esd1,esd2,esd3,esdc}. On the other hand, when some memory effect 
is present in the interaction between the system and the environment, 
i.e. when the noisy channel is non-Markovian \cite{Pii10,Sus10}, then 
a non monotone behaviour of entanglement and, more generally, of quantum 
correlations may be observed 
\cite{rev1,rev2,Man08,Maz09,Bel08,Har09,rev3,rev4,rev5}.  In fact,
entanglement oscillations are expected in continuous variable systems 
\cite{sab07,rug09}, whereas collapses and revivals of entanglement have 
been observed with polarization qubits \cite{Guo10}.
\par
In this Letter we suggest and demonstrate for the first time an
experimental setup to observe and engineer entanglement oscillations in
a programmable way. We address the spatial domain of 
spontaneous parametric downconversion, and exploit a
programmable spatial light modulator to impose a polarization- and
position-dependent phase-shift.  Since the polarization qubits are
obtained by tracing out the spatial degrees of freedom, our apparatus
allow us to analize the entanglement dynamics within the "coherence
time" of the effective non-Markovian channel. In this framework an
effective environment spectrum is obtained by acting on the spatial
profile of the SPDC. In turn, in order to investigate entanglement
oscillations we insert a spatial grating on the signal arm to achieve a
modulation of the environment spectrum.  Besides fundamental interest,
our scheme may found applications in engineering decoherence, e.g.  
in quantum process tomography.
\par
\begin{figure}[h!]
\includegraphics[width=0.82\columnwidth]{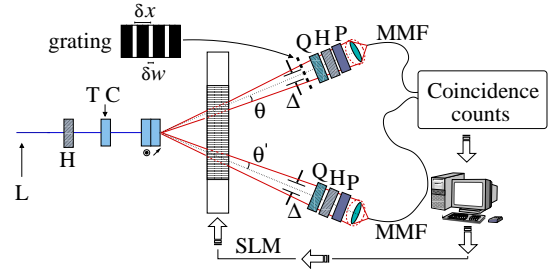}
\caption{(Color online) Schematic diagram of experimental setup. 
A linearly polarized cw laser diode at $406\,$nm (L) pumps
a couple of BBO crystals cut for Type-I downconversion. The horizontal a
nd vertical photon pairs are balanced by a half wave-plate set along the
pump path, whereas an additional BBO crystal is set on the pump path to
compensate the temporal delay. Signal and idler cones travel through the
SLM and are spatially selected by two irises and two slits set at
$D=500\,$mm with $\Delta x=5\,$mm ($\Delta=10\,$mrad). Two long-pass
filters cut-on wavelength $715\,$nm are used to reduce the background. A hand-made
grating can be inserted on the signal arm. Photons are focused in two multi-mode
fibers (MMF) and sent to single-photon counting modules. Polarizers at
the angles $45^\circ$, $-45^\circ$ or $45^\circ$, $45^\circ$ are inserted
to measure visibility whereas a quarter-wave plate, a half-wave plate and a
polarizer are used for the tomographic reconstruction.}\label{exp_setup}
\end{figure}

In our setup a two crystal geometry \cite{har92,Kwiat99,gen05} is 
used to produce two-qubit polarization entangled states by type-I 
downconversion in a non-collinear configuration. The state at the output 
of the crystal can be written as
\begin{align}
\ket{\psi}\!\propto\! \int\!\!\!\int\!\! d\theta d\theta' f(\theta,\theta')
&\left[ \ket{H\theta}\ket{H\theta'}+e^{\imath \Phi(\theta,\theta')}
\ket{V\theta}\ket{V\theta'}\right]\notag
\end{align}
where $\ket{P\theta}$ denotes a single photon state emitted
with polarization $P=H,V$ at angle $\theta$ ($\theta'$) along the
signal (idler) arm, $\theta$ and $\theta'$ are the shifts from the
central emission angle ($\theta_0,\theta'_0\simeq 3^\circ$), and
the integrations range from $-\frac12 \Delta$ to $\frac12
\Delta$, $\Delta$ being the angular aperture of two
slits along the downconversion paths, see Fig.~\ref{exp_setup}.
 \par
\begin{figure}[th!]
\includegraphics[width=0.7\columnwidth]{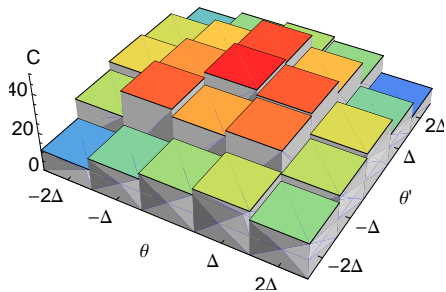}
\caption{(Color online) Coincidence counts distribution
$C=|f_{\textrm{exp}}(\theta,\theta')|^2$, within a coincidence
time window of $50$ns, as a function of the signal and
idler slit positions $\theta$ and $\theta'$. The phase matching central angles
correspond to $\theta, \theta'=0$.}\label{profili}
\end{figure}

The angle-dependent phase-shift $\Phi(\theta,\theta')=\phi(\theta)+
\phi'(\theta')+\Phi_0$ comes from the difference between the optical
path of vertically polarized photon pair, generated in the first
crystal, which must travel along the second one, and the optical path of
the pump beam traversing the first crystal before generating the
horizontally polarized pair in the second crystal. These angular
dependent terms are responsible for decoherence of polarization qubit
and should be removed in order to obtain an effective source of
entangled pairs \cite{kwi09}. Upon expanding to first order the terms in
$\Phi(\theta,\theta')$~\cite{cia10APL}, we arrive at
$\phi(\theta)=\gamma\theta$ and $\phi'(\theta')=-\gamma\theta'$. In our
apparatus, a one dimensional programmable spatial light modulator (SLM),
is set both on signal and idler path (see Fig. \ref{exp_setup}), and is
used to achieve a complete purification (i.e., $\Phi=\Phi_0$) by
inserting a linear phase function $\phi_\SLM(\theta)=-\gamma\theta$ on
the signal path and $\phi'_\SLM(\theta')=\gamma\theta'$ on the idler
path~\cite{guo09,cia10,cia10APL,cia08}. The constant phase $\Phi_0$ allows to
generate different maximally entangled states. In the present experiment
we set $\Phi_0=0$ by adding a proper constant phase to $\phi_\SLM$.  In
order to obtain an effective non-Markovian channel the SLM is then used
to impose an additional phase function $\phi_s(\theta)$ on the signal
arm.  Before going to details, let us devote some attention to the
angular function $f(\theta,\theta')$, which will be important for the
following discussion and assumed to have the factorized form
$g(\theta)g'(\theta')$. This assumption has been experimentally verified
by measuring the coincidence counts distribution
$C=|f_{\textrm{exp}}(\theta,\theta')|^2$, within a coincidence time
window of $50$ns, as a function of the signal and idler slit positions
$\theta$ and $\theta'$. We set two slits of aperture $\Delta x=1$mm
($\Delta=2$mrad) along the down-conversion arms and measured coincidence
counts within a time window of $10$s for slit positions
$\theta,\theta'=-2\Delta,-\Delta,\dots,+2\Delta$. In Fig.~\ref{profili}
we show the experimental data; the phase matching central angles
correspond to $\theta, \theta'=0$. The corresponding coincidence counts
distribution has then been compared with the one computed as
$\tfrac{\mid f_{\textrm{exp}}(\theta,0)
f_{\textrm{exp}}(0,\theta')\mid^2}{\mid f_{\textrm{exp}}(0,0)\mid^2}$
and an excellent agreement was found, also corroborated by a significant
$\chi^2$ test ($P_{\chi^2>\chi_0^2}\simeq 0.9$). From the results of
Fig.~\ref{profili} we also infer a Gaussian-like shape for the angular
distributions $g(\theta)$, $g'(\theta')$, with a FWHM of $8.6\,$mrad.
\par
Once the state has been purified and the additional phase
$\phi_s(\theta)=\alpha\theta$ has been imposed to the signal photon,
the dependence on $\theta'$ is traced out, and thus no terms
containing $g'(\theta')$ appear in the polarization
density matrix
$\varrho=\hbox{Tr}_{\theta,\theta'}\left[\ket{\psi}\bra{\psi}\right]
=\frac12 \left(
\ket{\HH}\bra{\HH}+\varepsilon\ket{\VV}\bra{\HH}+
\varepsilon^*\ket{\HH}\bra{\VV}+\ket{\VV}\bra{\VV}
\right)$
where $\varepsilon=\!\int\!d\theta \left| g(\theta)\right|^2 e^{\imath
\alpha \theta}$ is the decoherence factor.  It can be shown that, for
the state $\varrho$, the concurrence is $\mathcal C=|\varepsilon|$
\cite{Guo10}. Since the angular distribution $g(\theta)$ is symmetric,
$\varepsilon$ is real and positive, and we may write
$\varrho=\varepsilon \varrho_{b}+(1- \varepsilon) \varrho_{m}$, where
$\varrho_{b}$ denotes a Bell state and $\varrho_m$ the corresponding
mixture.  In turn, in this case, $\varepsilon$ equals the
interferometric visibility $V(\alpha)=\hbox{Re}[\varepsilon]$ which, in
turn, coincides with the concurrence $\mathcal C$.  Since we address the
spatial domain, it is straightforward to insert an amplitude modulation
on $g(\theta)$, e.g. by inserting a physical obstacle along the signal
optical path. Moreover, from the expression of the visibility
$V(\alpha)$, we see that a periodic structure of the angular
distribution would induce oscillations, whereas entanglement decrease
and then death may be expected for a non periodical angular
distribution.  In this framework $\alpha$ may be considered as the
evolution parameter of the dynamics of the (noisy) channel. In our
apparatus, the amplitude modulation is implemented by means of a
hand-made grating with a period $\delta x$ and a white region width
$\delta w$ centered along the signal arm (see Fig.~\ref{exp_setup}).  As
we will see, the narrower are the white regions the higher entanglement
oscillations are expected.  Formally, the insertion of the grating is
equivalent to the substitution $g(\theta)\rightarrow g(\theta)\cdot
m(\theta)$ in the visibility, up to the normalization $\int\! d\theta
\left| g(\theta) m(\theta)\right|^2=1$, where $m(\theta)$ is the
periodical unitary step function imposed by the grating. By simply
inserting or removing the grating it is possible to compare the
different dynamics imposed by a periodical or non-periodical angular
distribution.
\par
The experimental setup is shown in Fig.~(\ref{exp_setup}): a linearly
polarized cw, $406$nm laser diode (Thorlabs LPS-$406$-FC), with a
transverse profile $\textrm{TEM}_{00}$, pumps a couple of $1$mm thick
BBO crystals cut for Type-I downconversion. The $\ket{\HH}$ and
$\ket{\VV}$ pairs are balanced by a half wave-plate set along the pump
path. A BBO crystal with the proper length and optical axis angle is set
on the pump path, and is used to counteract the decoherence effect due
to the temporal delay between the two components
\cite{ser00,nam02,bar04,bri08,kwi09,cia10,cia10APL,cia08}.  This crystal
introduce a delay time between the horizontal and vertical polarization
of the pump which precompensates the delay time between the $\ket{\VV}$
pair generated by the first crystal and the $\ket{\HH}$ pair from the
second one.  Signal and idler cones travel through the SLM and are
spatially selected by two irises and two slits set at $D=500$mm. The low
quantum efficiency of our detectors ($\sim 10\%$) forces us to couple
large angular regions: we set $\Delta x=5$mm ($\Delta=10\,$mrad).
As we discuss in the following, this will decrease the maximum value of
the visibility. The down-conversion output is not spectrally filtered,
whereas two long-pass filters (cut-on wavelength $715$nm) are used to
reduce the background. A hand-made grating can be inserted on the signal
arm. Photons are focused in two multi-mode fibers (MMF) and sent to
home-made single-photon counting modules, based on an avalanche
photodiode operated in Geiger mode with passive quenching. In order to
measure the visibility, we insert two polarizers, set at the angles
$45^\circ$, $-45^\circ$ for the minimum and $45^\circ$, $45^\circ$ for
the maximum. For the tomographic reconstruction we insert on both paths
a quarter-wave plate, a half-wave plate and a polarizer.
\par
\begin{figure}[ht!]
\includegraphics[width=0.8\columnwidth]{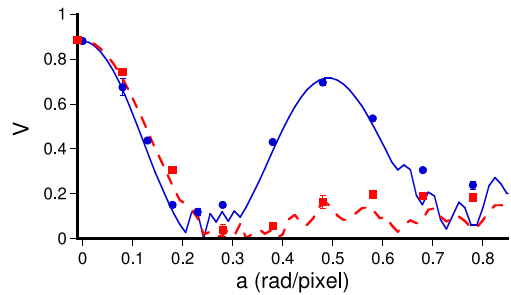}
\caption{(Color online) Visibility as a function of the evolution
parameter $a$. Blue circles and red square are the experimental data
obtained with and without grating (errors within the symbols). Blue
solid line and red dashed line denote the corresponding theoretical
predictions.
}\label{curva_revival}
\end{figure}

After purification we study the behaviour of the visibility as a
function of the dimensionless evolution parameter $\alpha$, governing
the linear phase function $\phi_s(\theta)=\alpha \theta$ imposed to the
signal by the SLM.  As previously discussed, oscillations of
entanglement are expected when the grating is inserted. Because of the
pixel discretization a step-function with an angular resolution
$\zeta=0.3\,$mrad is physically inserted by the SLM in order to
approximate the linear functions $\phi_{\SLM}(\theta),\phi'_{\SLM}(\theta')$.  
Experimentally, using the SLM, we impose the phase functions
\begin{align}
\phi_{\SLM}^e(n)&=-a_{opt} n + b \qquad\quad\:\hbox{on idler}\notag \\
\phi_{\SLM}^e(n)&=a_{opt} n  +\phi_s^e(n) \qquad\hbox{on signal}
\notag\,,
\end{align}
where $n$ is the distance in pixels from the center of the signal beam
($n=0$ for $\theta=0$), $a_{opt}=0.12\,$rad/pixel is the optimal slope
which allows us to achieve a complete purification, and $b=-\Phi_0$. The
linear function $\phi_s^e(n) = a\, n$ is also inserted to study the
dynamics, where the experimental evolution parameter is given by
$a=\alpha h/L\,$rad/pixel, $h=100\mu$m being the pixel width and
$L\simeq330$mm the distance between the SLM and the generating crystals.
Since the pixel discretization of the SLM imposes the condition
$a\ll2\pi/\textrm{pixel}$, high values of $a$ must be neglected in our
analysis. We experimentally verified that the curve $V(a)$ saturates to
the uncompensated value when $(a+a_{opt})\to 2\pi$.  The revival is
expected at $a_{rev}=\tfrac{2\pi D}{\delta x}\tfrac{h}{L}$ or, in terms
of the angular grating period $\delta\theta$,
$\alpha_{rev}=\tfrac{2\pi}{\delta\theta}$. We choose $\delta x=2$mm,
which leads at $a_{rev}=0.476\,$rad/pixel, in order to avoid high values
of the evolution parameter, and we set $\delta w= 0.4 \delta
x$.  In Fig.~(\ref{curva_revival}) we present the experimental results,
together with the theoretical prediction calculated from the expression
of the visibility, as a function of the experimental evolution parameter
$a$.  Blue circles and red squares are the experimental data obtained
with and without grating, respectively. Blue solid line and red dashed
line are the theoretical predictions.
\par
\begin{figure}[ht!]
\includegraphics[width=0.35\columnwidth]{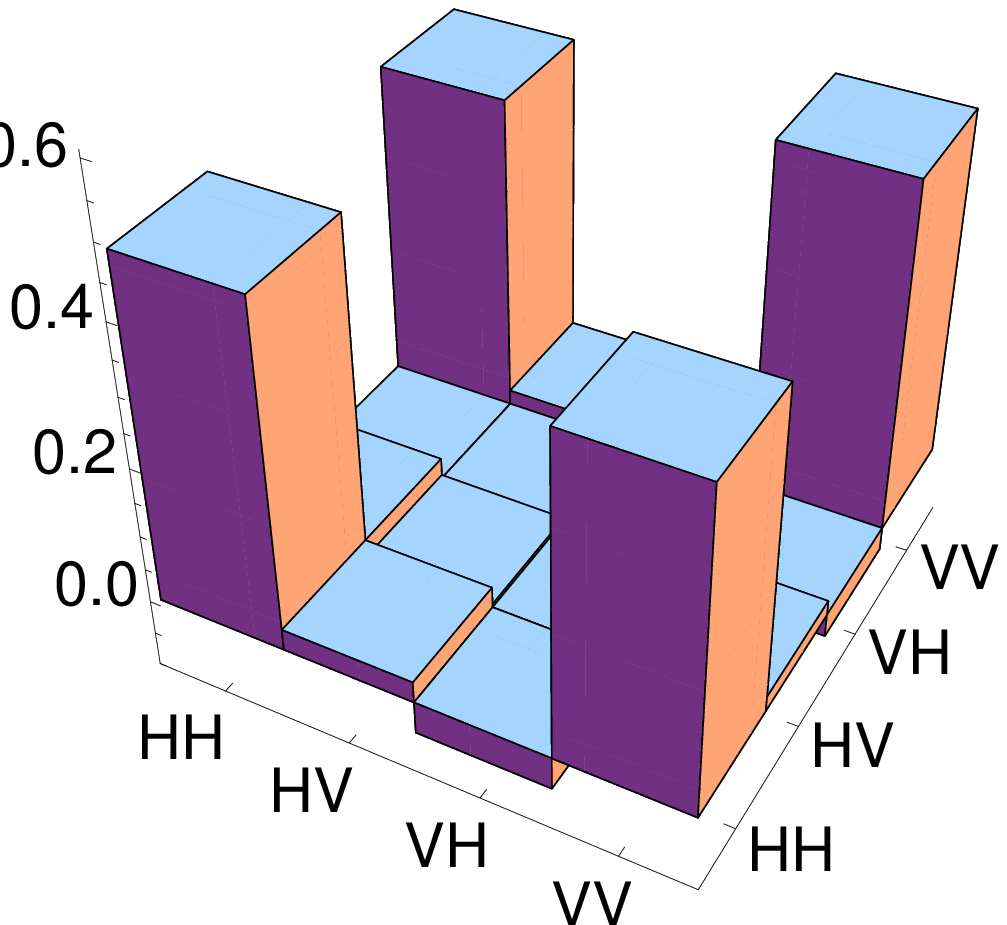}
\includegraphics[width=0.35\columnwidth]{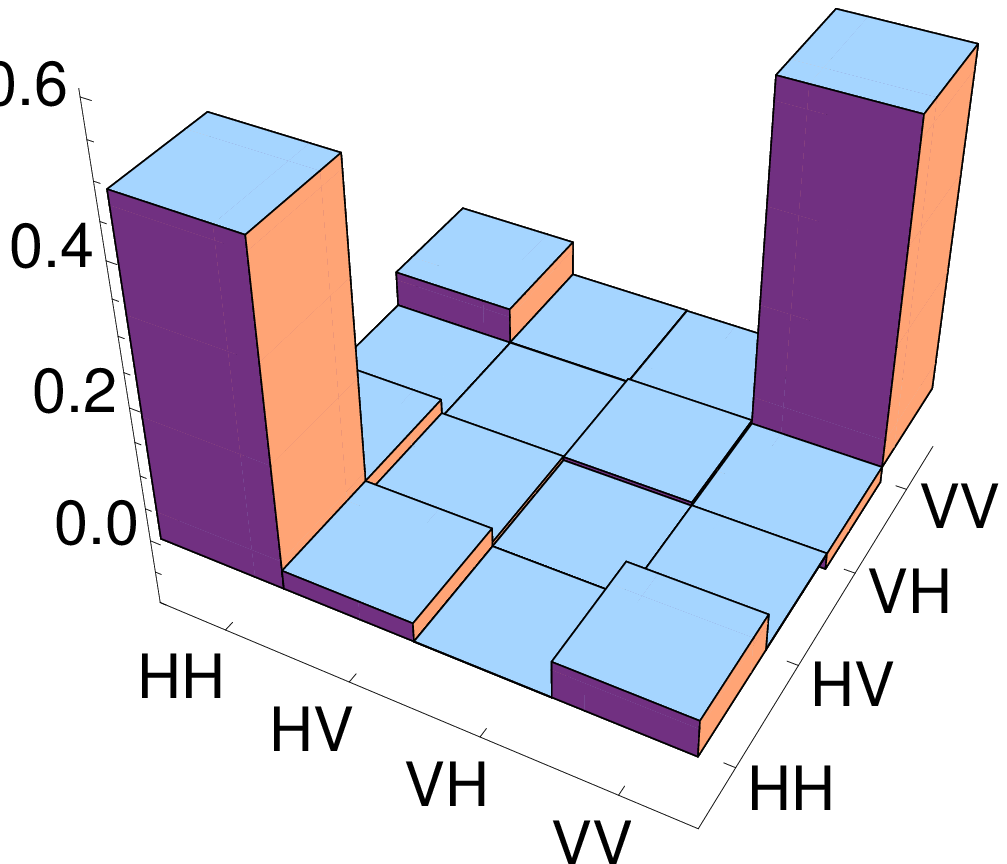}
\includegraphics[width=0.35\columnwidth]{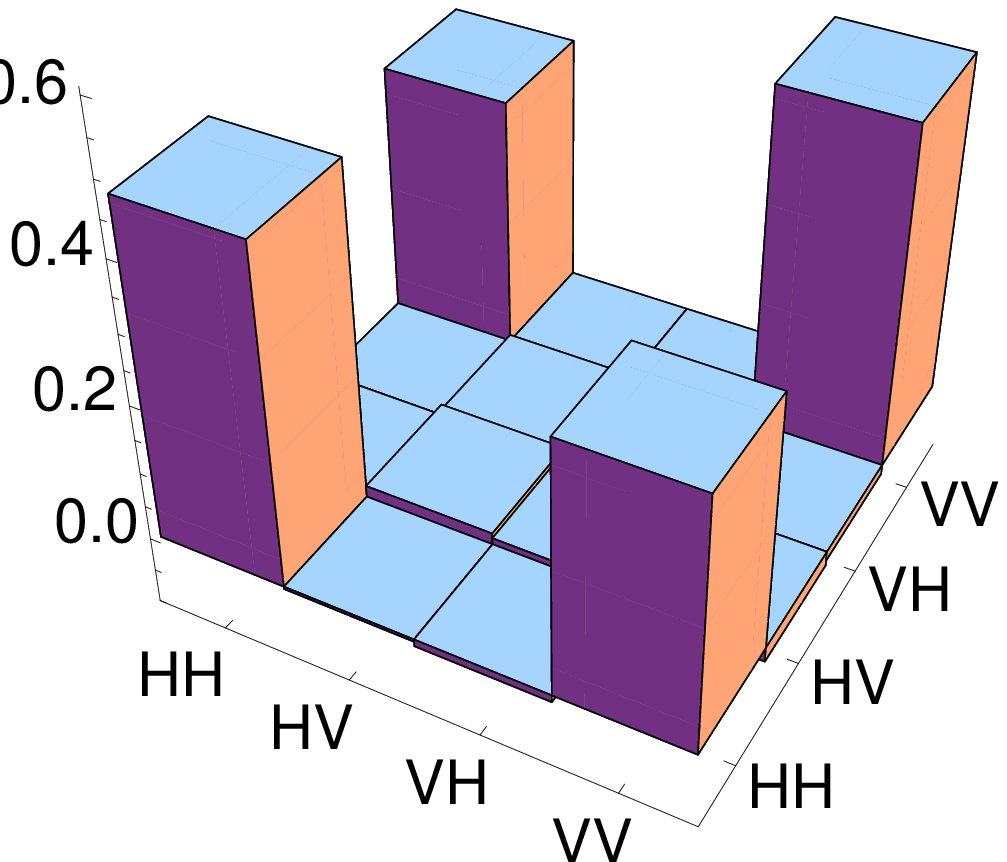}
\includegraphics[width=0.35\columnwidth]{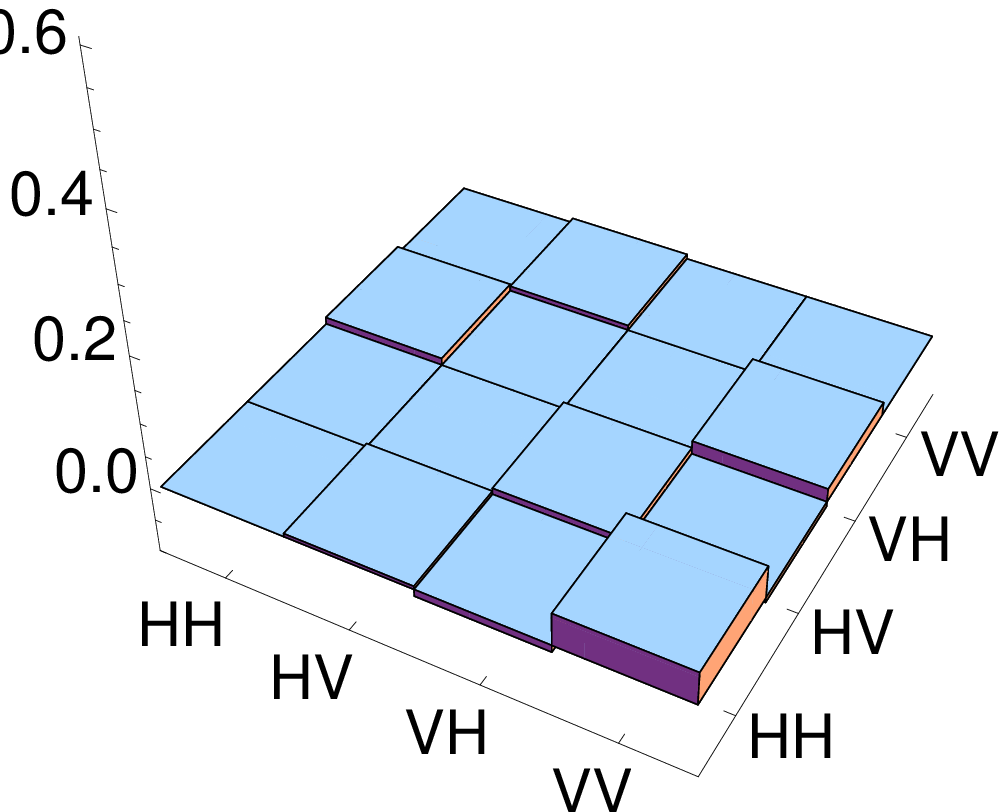}
\caption{(Color online) Tomographic reconstruction of a state 
evolving in the effective non-Markovian channel. In the upper left
plot the two-qubit density matrix just after the purification, with 
visibility $V=0.881 \pm 0.004$. Upon increasing the
evolution parameter to $a=0.23$ we achieve the minimum of
entanglement oscillations: the density matrix is
shown in the upper right plot, the corresponding visibility
is $V=0.120\pm 0.016$. In the lower panels,
we show the real and the imaginary part of the reconstructed
density matrix at the maximum of entanglement oscillations, which
occurs for $a=0.48$. The corresponding
visibility is $V=0.696\pm0.013$, and the Bell parameter $B=2.341\pm0.019$.
}\label{qtomo}
\end{figure}

In order to fully characterize the output state we have also performed
state reconstruction by polarization qubit tomography for different
values of the evolution parameter $a$. The procedure goes as follows: we
measure a suitable set of independent two-qubit
projectors~\cite{mlik00,jam01} and then reconstruct the density matrix
from the experimental probabilities using maximum-likelihood
reconstruction of two-qubit states. The tomographic measurements are
obtained by inserting a quarter-wave plate, a half-wave plate and a
polarizer.  The purification procedure with the grating inserted leads
to a visibility $V=0.881 \pm 0.004$, the density matrix is graphically
represented in the upper left panel of Fig.~\ref{qtomo}.  Increasing the
evolution parameter to $a=0.23 $, the visibility decreases to
$V=0.120\pm 0.016$. The corresponding tomographic reconstruction,
depicted in the upper right panel of Fig.~\ref{qtomo}, well illustrate
the degradation of entanglement. However, we found an relevant revival
after a further increasing of the evolution parameter to $a=0.48$, where
we have $V=0.696\pm 0.013$. The corresponding tomographic reconstruction
(real and imaginary parts) are reported in the lower panels of
Fig.~\ref{qtomo}.  In order to show the revival of the nonlocal
correlations we have also measured  the Bell parameter
$B=\left|E(\beta_1,\beta_2) + E(\beta_1,\beta_2^\prime) +
E(\beta_1^\prime,\beta_2)- E(\beta_1^\prime,\beta_2^\prime) \right|$
where $E(\beta_1,\beta_2)$ denotes the correlations between measurements
performed at polarization angle $\beta_j$ for the mode $j$. We found
$B=2.341\pm0.019$, which violates CHSH-Bell  inequality~\cite{CHSH} by
more than $17$ standard deviations.  Comparing this curve with the one
obtained without the grating we see that in the latter case no revival
occurs after the degradation of entanglement. We also notice that the
minimum occurs for lower values of the evolution parameter compared to
the case with the grating.
\par
The residual lack of visibility after the purification procedure is
mainly due to the very broad downconversion spectral range that is
selected by the slits. In fact, with the selected slit aperture, $\Delta
x = 5\,$ mm $\rightarrow \Delta=10\,$mrad, we estimate that about
$200\,$nm of the downconversion spectrum are coupled. By narrowing the
slit aperture to $\Delta=4\,$mrad only $60\,$nm are selected, and
the visibility is found to increase at $V=0.963\pm0.005$. This suggests
some achromatic effect in the action of the SLM, which prevents a perfect
purification.  The present experiment has been performed with the larger
aperture to compensate the low quantum efficiency of photodetectors.
\par
In conclusion, we have suggested and demonstrated an experimental setup
to observe oscillations of polarization entanglement in a programmable
way.  Our scheme is based on a spatial light modulator, which is
inserted on the spatial domain of the downconversion output to achieve
an effective non-Markovian environment, and on a grating, which has been
employed to create a tunable environment spectrum. Our scheme is
all-optical and allows us to generate and detect revivals of
entanglement and nonlocality, thus paving the way for engineering of
decoherence for polarization qubits.
\par
MGAP thanks Sabrina Maniscalco, Stefano Olivares, Ruggero Vasile, 
Pino Vallone, Marco Genovese, and Bassano Vacchini for useful discussions.

\end{document}